\documentclass{aa}

\usepackage{lscape}
\usepackage{amsfonts,amsmath,natbib,xcolor,subfigure,gensymb,longtable,ulem}
\usepackage{xtab,booktabs}
\usepackage{hyperref}
\hypersetup{breaklinks,colorlinks,citecolor=blue,linkcolor=blue,urlcolor=red}

\usepackage{graphicx}
\usepackage{txfonts}

\usepackage{orcidlink}  % Place ORCID links

\begin{document}

\title{A search for periodic activity in multi-peaked long gamma-ray bursts}
\author{C.~Guidorzi\thanks{guidorzi@fe.infn.it}\inst{\ref{unife},\ref{infnfe},\ref{inafbo}}\orcidlink{0000-0001-6869-0835}
\and M.~Orlandini\inst{\ref{inafbo}}\orcidlink{0000-0003-0946-3151}
\and R.~Maccary\inst{\ref{unife},\ref{inafbo}}\orcidlink{0000-0002-8799-2510}
\and L.~Amati\inst{\ref{inafbo}}\orcidlink{0000-0001-5355-7388}
\and L.~Bazzanini\inst{\ref{unife},\ref{inafbo}}\orcidlink{0000-0003-0727-0137}
\and M.~Bulla\inst{\ref{unife},\ref{infnfe},\ref{inafte}}\orcidlink{0000-0002-8255-5127}
\and L.~Ferro\inst{\ref{unife},\ref{inafbo}}\orcidlink{0009-0006-1140-6913}
\and F.~Frontera\inst{\ref{unife},\ref{inafbo}}\orcidlink{0000-0003-2284-571X}
\and M.~Maistrello\inst{\ref{unife},\ref{inafbo}}\orcidlink{0009-0000-4422-4151}
\and W.-J.~Tan\inst{\ref{ihep},\ref{ucas}}
\and A.~Tsvetkova\inst{\ref{unica},\ref{ioffe}}\orcidlink{0000-0003-0292-6221}
\and S.~L.~Xiong\inst{\ref{ihep}}\orcidlink{0000-0002-4771-7653}
\and S.~N.~Zhang\inst{\ref{ihep}}\orcidlink{0000-0001-5586-1017}
}

\institute{Department of Physics and Earth Science, University of Ferrara, Via Saragat 1, I-44122 Ferrara, Italy \label{unife}
\and  INFN -- Sezione di Ferrara, Via Saragat 1, 44122 Ferrara, Italy\label{infnfe}
\and INAF -- Osservatorio di Astrofisica e Scienza dello Spazio di Bologna, Via Piero Gobetti 101, 40129 Bologna, Italy \label{inafbo}
\and INAF, Osservatorio Astronomico d'Abruzzo, via Mentore Maggini snc, 64100 Teramo, Italy \label{inafte}
\and Key Laboratory of Particle Astrophysics, Institute of High Energy Physics, Chinese Academy of Sciences, Beijing 100049, People's Republic of China \label{ihep}
\and University of Chinese Academy of Sciences, Chinese Academy of Sciences, Beijing 100049, People's Republic of China\label{ucas}
\and Department of Physics, University of Cagliari, SP Monserrato-Sestu, km $0.7$, 09042 Monserrato, Italy \label{unica}
\and Ioffe Institute, Politekhnicheskaya 26, 194021 St. Petersburg, Russia \label{ioffe}
}

\abstract
{A sizeable fraction of gamma-ray burst (GRB) light curves (LCs) feature a sequence of peaks that hold information on the unknown way energy is released and dissipated into gamma-rays over time. Traditionally, searches for periodic signals in GRB LCs have turned out to be inconclusive, partly because they are challenging as a consequence of the short-lived, coloured-noise, and non-stationary nature of the LCs themselves. Yet, recent claims have revived the issue.}
{We aim to search for periodic components in GRB LCs through a new approach to GRBs that avoids most of the issues faced by traditional Fourier techniques and periodograms due to the GRB LC nature.}
{We identified peaks through a well-tested algorithm and selected GRBs with at least ten peaks out of five GRB catalogues ({\it Swift}/BAT, CGRO/BATSE, {\it Fermi}/GBM, Insight-HXMT, and {\it Beppo}SAX/GRBM). At this point, each GRB was simply treated as a discrete point process whose realisation coincides with the sequence of peak times. We searched for possible periodic recurrences based on the multinomial distribution, after accounting for the clustering of peaks due to the non-stationarity of the GRB signals.}
{Accounting for the multiple trial periods, the best candidate has a p value of $3\times10^{-4}$ that there is no periodic recurrence. However, accounting for the multiple trials of 555 searched GRBs, its statistical significance is downgraded to 17\%. The overall distribution of the p values obtained for all GRBs is compatible with a uniform distribution in $[0,1]$. Our technique is sensitive to the presence of $\gtrsim 8$ periodic peaks and a periodic peak fraction of $\sim75$\% at least.}
{We found no robust evidence for multi-peaked GRBs with periodic recurrences at peak times. We can exclude a sizeable fraction ($\gtrsim 0.75$) of peaks of each GRB with at least ten peaks being periodic. While our result does not necessarily clash with claimed periodicities based on Fourier techniques, it constrains the putative recurrent behaviour, which would manifest itself not through the sequence of peaks but, evidently, in a more elusive way.}

\maketitle
%%%%%%%%%%%%%%%%%%%%%%%%%

%%%%%%%%%%%%%%%%%%%%%%%%%%%%%%%%%%%%%%%%%%
\section{Introduction}
\label{sec:intro}
%%%%%%%%%%%%%%%%%%%%%%%%%%%%%%%%%%%%%%%%%%
The origin of the prompt emission of gamma-ray bursts (GRBs) remains one of the most debated and open issues in GRB physics.
A number of alternative scenarios have been put forward to explain three main intertwined aspects: (i) the uncertain nature of the newborn compact object; (ii) the process through which this inner engine powers the relativistic outflow; and (iii) the mechanism that dissipates a fraction of energy into gamma-rays (see \citealt{Zhang18_book} for a comprehensive review).

The possibility of identifying a periodic component would provide a strong direct clue about (i) and (ii), and indirectly about (iii), too.
Searching for periodic or quasi-periodic oscillations (QPOs) can be carried out either in frequency or in the time domain. In frequency, the presence of red noise in the power density spectrum (PDS) is known to alter the statistical properties of power from a simple $\chi^2$ distribution, which is instead expected in the presence of white noise (see \citealt{Vanderklis88} for a review). As a consequence, searching for periodicity in the presence of red noise is more challenging (e.g. \citealt{Israel96,Vaughan10} and references therein).
Moreover, when we look for the statistical properties of a random process generating the observed time series, we tacitly assume that a single realisation of the process provides sufficient information to infer the overall statistical properties of the process. In other words, we assume that the stochastic process is `ergodic'.  For this to happen, the process must be stationary, in the sense that all the moments and joint moments of the random process do not depend on the particular time at which they are computed. The timing analysis of non-ergodic processes can therefore produce incorrect results, such as highly overestimating the presence and significance of putative QPOs  \citep{Huebner22a}, and significantly affect the PDS because of red noise.

In this respect, GRB light curves (LCs) are very challenging, since they are short-lived, have individual and average PDSs that are significantly affected by red noise \citep{Beloborodov98,Guidorzi12,Dichiara13a,Guidorzi16}, are overall non-stationary (e.g. \citealt{SternSvensson96,Bazzanini24}, being the result of an irreversible catastrophic process), and, as an ensemble, display a great variety of morphologies and characteristic timescales.

In the literature, a number of searches for periodicity in GRB LCs, either systematic or focused on some cases of special interest, have appeared so far. One can divide them based on the frequency range: (a) high-frequency ($10$--$10^3$~Hz) and (b) low-frequency ($0.01$--$10$~Hz).

Concerning the low-frequency range, no unambiguously convincing case ($>3\sigma$ significance) has been reported to date, although different conclusions based on wavelets have occasionally been reported \citep{Tarnopolski21}. In spite of initial claims, the putative oscillations found in GRB\,090709A with a period of $8.1$~s were soon downgraded, upon careful modelling of the red noise part of the PDS \citep{DeLuca10,Cenko10}. A similar conclusion was reached recently through a Bayesian inference approach in the time domain, fitting the LC of GRB\,090709A with a deterministic plus a stochastic component modelled in terms of a stationary Gaussian process \citep{Huebner22b}.
No convincing case has emerged from systematic analyses of samples of long GRBs that has properly accounted for the red noise affecting their PDSs, especially in the low-frequency domain ($\lesssim 1$~Hz) and for the multi-trial that is implied in terms of both the number of examined frequencies and the number of analysed GRBs \citep{Guidorzi16,Xiao22}.

In the high-frequency domain, recent claims of kilohertz periodic components in a couple of short bursts detected with the Burst And Transient Source Experiment (BATSE) have been reported \citep{Chirenti23}; however, they have not been confirmed by independent analyses \citep{LiuZou24}. A similar claim was reported for GRB\,200415A \citep{Castro-Tirado21}, which, despite its name, is likely a giant flare of an extragalactic magnetar \citep{Roberts21,Svinkin21,LAT21}, belonging therefore to a class of sources that, in the Galaxy, have occasionally exhibited high-frequency QPOs in the tail of analogous events \citep{Israel05,StrohmayerWatts05}. Other similar claims, although at significantly lower frequencies, have recently appeared: in particular, a $\sim$20-Hz oscillatory component in the long-duration merger candidate GRB\,211211A \citep{Xiao24,Chirenti24}. According to these analyses, these transient oscillations are observed in a $0.2$-s long interval. Within such a short window, a $\sim$20-Hz transient oscillation can only accomplish four cycles, which corresponds to a very low quality factor, $Q$\footnote{In Fourier space, it is $Q=f/\Delta f$, where $f$ and $\Delta f$ are the central frequency and width, respectively, for a QPO. Within a window, $T$, there are $N = f T$ oscillations. The QPO cannot be narrower than the frequency resolution: $\Delta f \ge 1/T$, which implies $Q\le N$.}, notwithstanding that, being so close to the low-frequency boundary of the explorable range, it requires an accurate modelling of the red noise, which can hardly be achieved.

If confirmed, these QPOs would provide strong clues about how these binary compact object mergers happen, possibly related to oscillations experienced by the unstable and strongly magnetised hyper-massive neutron star being formed \citep{MostQuataert24,Xiao24}.
Yet, the evidence so far used to support the statistical significance of these QPOs crucially relies on the assumptions that the null-hypothesis scenario of no QPO is built upon. In this sense, non-stationarity may undermine the conclusions based on assuming stationarity; in any case, it makes the choice of time intervals to search for QPOs crucial and somewhat arbitrary. Moreover, modelling using Gaussian processes inevitably restricts in an arbitrary way the possible stochastic nature of the GRB signal; for example, should the statistical noise\footnote{This is an intrinsic property of the signal and is not to be confused with the counting statistics noise due to the measurement process.} be multiplicative rather than additive -- and we see no reason why one should exclude this scenario a priori -- the stochastic component could not be described as a Gaussian process, either invalidating or undermining the QPO claims. Or, still assuming the validity of a Gaussian process description, one cannot use integration times that are too short, since the low-count regime would require a Poisson process. In addition, how all the multi-trial aspects are accounted for is often overlooked. How many time intervals have been searched prior to the interesting ones being identified, and how many GRBs have been considered before the promising ones are focused on, are just a few subtle but potentially key issues that one should consider before making claims about (quasi-) periodic oscillations in transient and non-stationary events such as GRBs. 

In this work, we try a different approach that has never been applied to GRBs to our knowledge: we identify peaks in long GRBs with many peaks (at least ten) through {\sc mepsa} \citep{Guidorzi15a}, a well-tested algorithm, and turn the GRB LC into a temporal sequence of events, given by the peak times. Consequently, the problem of searching for oscillatory components turns into a problem of searching for periodic recurrences in the peak times, which is in principle much simpler to address. 

A peak is identified by {\sc mepsa} whenever the rate of a given bin exceeds those of its surrounding left and right bins ($n_l$ and $n_r$, respectively) by a corresponding sequence of values in units of $\sigma$ (that is, combined rate uncertainties): several sequences, called  ``patterns'', are tried and a peak is identified when at least one of these patterns is fulfilled. This scheme is repeated by rebinning the LC by all values within a range and, for each rebinning factor, the different LCs with all possible initial phases are examined.
The advantage of using {\sc mepsa} mainly lies in its capability to detect, across different timescales, several peak structures that are observed in GRB LCs, since it was explicitly calibrated with that purpose, while keeping a very low rate of false positives, unlike previous analogous algorithms. It has two main limitations: (i) the noise is assumed to be Gaussian: when this is barely the case, as in the Poisson low-count regime, one may still use it, provided that a conveniently high threshold on the signal-to-noise ratio (S/N) of the peak candidates is adopted; (ii) compared with an experienced GRB-trained human eye, {\sc mepsa} seems to miss about 10\% of all peaks, especially when they overlap enough to be missed by the code but not by the well-trained eye \citep{Maccary24}.

Having transformed the GRB time series into a point process has essentially removed issues such as red noise. Non-stationarity remains a key issue, since it is responsible for the clustering of peaks with time, and we show how this was addressed in the new reformulation of the problem. While the basic idea is not new, and in fact was used to discover the periodic recurrence of bursts of some of the repeating fast radio burst (FRB) sources \citep{CHIME20a}, applying it to GRBs is different and somehow more challenging, due to the combination of short-livedness and non-stationarity. 

Section~\ref{sec:data} describes the data sets. Section~\ref{sec:ana} illustrates the technique adopted to search for periodic recurrences of peaks along with its calibration. The results are presented in Sect.~\ref{sec:res} and discussed in Sect.~\ref{sec:disc_conc}, in which conclusions are finally drawn.

%%%%%%%%%%%%%%%%%%%%%%%%%%%%%%%%%%%%%%%%%%
\section{Data sets}
\label{sec:data}
%%%%%%%%%%%%%%%%%%%%%%%%%%%%%%%%%%%%%%%%%%
%--------------------------------
\subsection{{\it Swift}/BAT sample}
\label{sec:swift}
%--------------------------------
Starting from the long GRBs detected by {\it Swift}/BAT in burst mode from January 2005 to October 2024, we extracted the mask-weighted LCs with a uniform bin time of 4~ms in the 15-150~keV energy passband, following the prescriptions of the BAT team\footnote{\url{https://swift.gsfc.nasa.gov/analysis/threads/bat_threads.html}.}.
Peaks were identified through {\sc mepsa}, imposing a S/N threshold of S/N $\ge 5$. We then selected the GRBs with a minimum number of $N_p=10$ peaks, ending up with a sample of 80 GRBs that is hereafter referred to as the {\it Swift} sample.

%--------------------------------
\subsection{CGRO/BATSE sample}
\label{sec:batse}
%--------------------------------
BATSE did not cover long-lasting GRBs in TTE entirely because of on-board memory limitations. The best temporal resolution available for the whole profiles of long GRBs is 64~ms. We therefore used the results of the peaks identified for each BATSE GRB with {\sc mepsa}, which are reported in \citet{Guidorzi24}. These LCs were taken with the eight BATSE Large Area Detectors (LADs) and were obtained by the BATSE team from the concatenation of three standard BATSE types, DISCLA, PREB, and DISCSC, in four energy channels: 25-55,
55-110, 110-320, and $>320$~keV. We used the summed counts of these four channels. We interpolated the background with polynomials of up to the fourth degree, as was prescribed by the BATSE team.

The BATSE catalogue has the best sensitivity and the largest average number of peaks per GRB among all past and present GRB experiments. The same selection criteria, S/N $\ge 5$ and  $N_p \ge 10$, were applied. As a consequence, we ended up with a sample of 209 GRBs that is hereafter referred to as the BATSE sample.

%--------------------------------
\subsection{{\it Fermi}/GBM sample}
\label{sec:fermi}
%--------------------------------
From the catalogue of long GRBs provided by the {\it Fermi} team, we selected all the long $(T_{90}>2$~s) GRBs from July 14 2008 to February 4 2024. We ignored the few very bright GRBs that saturated the GBM detectors, such as GRB\,130427A and GRB\,221009A \citep{Preece14,Lesage23}. We rejected the GRBs that are affected by the simultaneous occurrence of a solar flare or whose profile was not entirely covered by the time-tagged event (TTE) mode of GBM.

For each GRB, we extracted the LCs of the most illuminated NaI detectors in the 8--900~keV energy passband with a uniform bin time of 4~ms.
The background was interpolated and subtracted using the GBM data tools\footnote{\url{https://fermi.gsfc.nasa.gov/ssc/data/analysis/gbm/gbm_data_tools/gdt-docs/}.}
\citep{GbmDataTools} following standard prescriptions (see \citealt{Maccary24} for details). The quality of the background interpolation was assessed by ensuring that the normalised residuals in the background intervals were compatible with uncorrelated Gaussian noise, as is explained in \citet{Maccary24b}.
For each GRB, the selection of the GBM units was based on the `scat detector mask' entry on the HEASARC catalogue\footnote{\url{https://heasarc.gsfc.nasa.gov/db-perl/W3Browse/w3table.pl?tablehead=name\%3Dfermigbrst&Action=More+Options}}. We used the TTE data from the start of its $T_{90}$ interval to the end.
Finally, we identified and removed the charged particle spikes from the LCs as follows: bins whose counts exceeded by $\ge 9\sigma$ the counts of the adjacent bins were tagged as being due to a potential spike. Whenever visual inspection of the different detectors established the spurious nature of the counts within a given bin, as was revealed by a set of peak intensities that were incompatible with being caused by a plane electromagnetic wave, its counts were replaced with the mean of the adjacent bins.

We finally applied {\sc mepsa} to the background-subtracted LCs, imposing the same threshold of S/N $\ge 5$ and at least ten peaks per GRB as for the {\it Swift} sample. In this way, we came up with a sample of 240~GRBs from {\it Fermi}/GBM, which will hereafter be referred to as the GBM sample.

%--------------------------------
\subsection{Insight-HXMT sample}
\label{sec:hxmt}
%--------------------------------
The first GRB catalogue of  Insight-HXMT \citep{Song22_HXMTGRBcatalog} includes GRBs that were observed with the High Energy instrument (HE; \citealt{Liu20_HXMT}) used as an open-sky monitor. From this catalogue, we selected the long-duration ($T_{90}>2$~s) GRBs and extracted their 4-ms background-subtracted LCs, summing up the counts of all the 18 HE detectors in the nominal band (either 80--800 or 200--3000~keV,\footnote{These passbands refer to the deposited energies of incident photons.} depending on whether the operation mode of the HE instrument at the time of each GRB was normal or low-gain mode, respectively). Background interpolation was carried out as in \citet{Camisasca23}.

We applied {\sc mepsa} to all background-subtracted LCs and considered the peaks with S/N $\ge 5$,\footnote{In \citet{Camisasca23} we applied a higher, more conservative threshold on the peaks detected at 4-ms timescale, since in that case the purpose was determining the minimum variability timescale. Consequently, very short peaks had to be very robust.} in line with the procedures applied to the other experiments. We then selected the GRBs by imposing $N_p\ge 10$ on the number of peaks. Excluding the very intense GRBs that temporarily saturated the electronics (the so-called `iron' sample in \citealt{Song22_HXMTGRBcatalog}; see also \citealt{Camisasca23}), we finally ended up with nine GRBs in what is hereafter referred as the HXMT sample.

%--------------------------------
\subsection{{\it Beppo}SAX sample}
\label{sec:bepposax}
%--------------------------------
The GRB catalogue of the {\it Beppo}SAX Gamma-Ray Burst Monitor (GRBM; \citealt{Frontera97}) presents two kinds of temporal resolutions, depending on whether any given GRB triggered the on-board logic: 1~s (ratemeters) and the high-time resolution of $7.8125$~ms lasting 106~s, with the latter mode available only for the on-board triggered events, which make up $\sim 2/3$ of the whole catalogue \citep{Frontera09}. The energy passband is 40--700~keV. We focused on the long ($T_{90}>2$~s) GRBs that triggered the on-board logic and applied {\sc mepsa} to the $7.8125$-ms background-subtracted LCs that have the best S/N: the GRBM had four independent units that formed a square well. In most cases, the LC with the best S/N was given by the sum of the two most illuminated units and only rarely by just one unit. Background interpolation and particle spike removal were obtained in the manner described in \citet{Frontera09}. We then selected all GRBs that have $N_p\ge 10$ peaks with S/N $\ge 5$, ending up with a sample of 17 GRBs that is hereafter referred to as the {\it Beppo}SAX sample. The 555 GRBs from the five samples form together what is referred to as the total sample.

%%%%%%%%%%%%%%%%%%%%%%%%%%%%%%%%%%%%%%%%%%
\section{Data analysis}
\label{sec:ana}
\begin{figure}
   \includegraphics[width=0.47\textwidth]{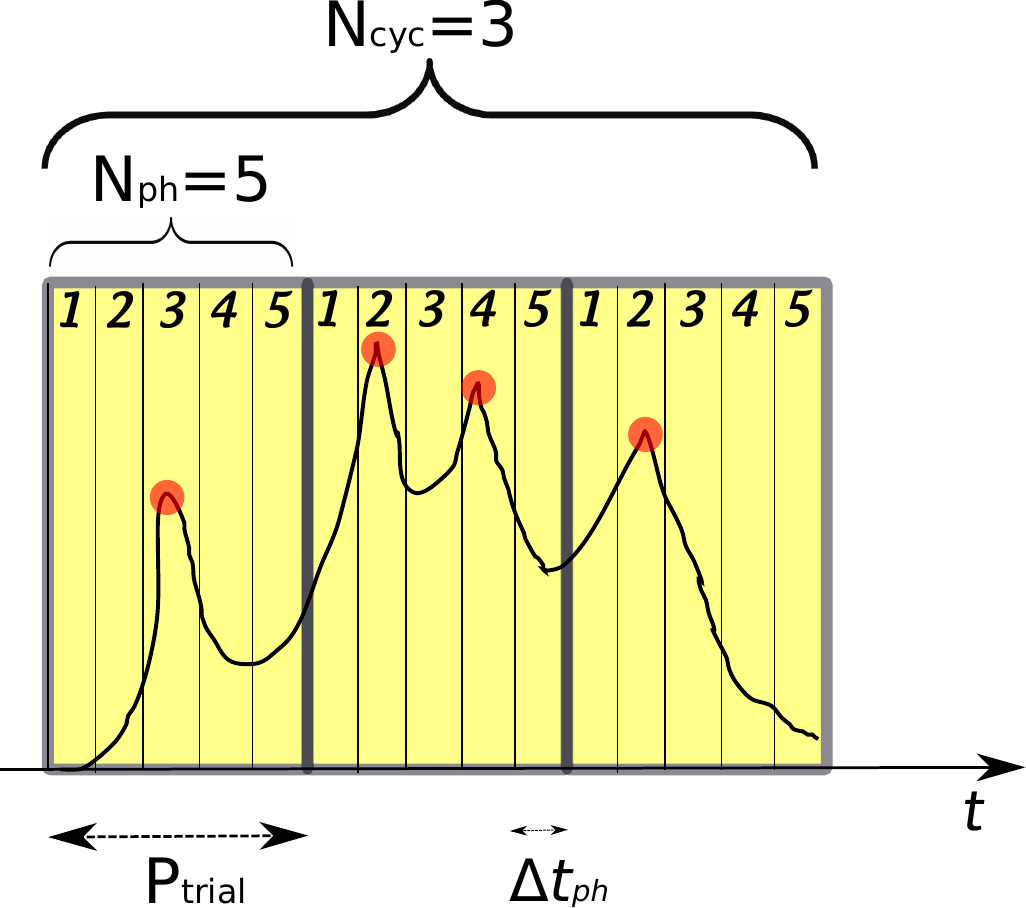}
    \caption{Illustrative sketch of how the folding and the related search for periodic occurrences of peaks was carried out.}
    \label{fig:drawing}
\end{figure}

%%%%%%%%%%%%%%%%%%%%%%%%%%%%%%%%%%%%%%%%%%
The key idea is to fold the time window that includes a given GRB according to a number of trial periods, dividing each period into a number of phase bins, and see whether the peak times tend to cluster within one or a few phase bins: when this is the case and is statistically significant, this would imply that peaks do not occur randomly, but repeat with a period that is commensurate\footnote{Two quantities are said to be commensurate when their ratio equals the ratio between two integers.} with the trial one.
Figure~\ref{fig:drawing} illustrates the essence of the technique and the meaning of some of the quantities defined in the following, along with a sketch of a GRB LC, with its peaks marked with red circles.
The problem of evaluating the likelihood of a given distribution of $N_p$ peak times within $N_{\rm ph}$ different phase bins is formally equivalent to having $N_p$ balls within $N_{\rm ph}$ bowls: this is ruled by the multinomial distribution.

We let $t_{p,i}$ be the time of the $i$-th peak and $\delta t_{p,i}$ its detection timescale assigned by {\sc mepsa}, which can be taken as the uncertainty on $t_{p,i}$. We first defined the time interval that includes the whole GRB as $[t_i, t_f]$, where $t_i = t_{p,1}-\tau$, $t_f = t_{p,N_p} + \tau$, where $\tau = (t_{p,N_p}-t_{p,1})/10$. In other words, we took the time interval from the first to the last peak time and expanded on each side by one tenth. We let $T = t_f-t_i$ be the duration of the whole time interval.

Here, we define a few key parameters:
\begin{itemize}
    \item $N_{\rm cyc}$ is the number of cycles that a putative periodic signal would have within $T$; it ranges between $N_{\rm cyc,min}=8$ and $N_{\rm cyc,max}$, which varies for different GRBs.
    \item $P$ is the corresponding trial period and is $P = T/N_{\rm cyc}$; it ranges between $P_{\rm min} = T/N_{\rm cyc,max}$ and $P_{\rm max} = T/N_{\rm cyc,min}$.
    \item $N_{\rm ph}$ is the number of phase bins, into which a time interval corresponding to a trial period is divided; it ranges between $N_{\rm ph,min}=5$ and $N_{\rm ph,max}$, which depends on $P$. We thoroughly examined the alternative option of fixing this parameter to a single and optimally chosen value for each GRB. As a result, relatively large values of $N_{\rm ph}$ performed equally well in terms of sensitivity, but at the cost of remarkably shrinking the range of explorable periods. We therefore opted for the variable $N_{\rm ph}$ solution.
    \item $\Delta t_{\rm ph}$ is the phase bin duration, which is given by $P/N_{\rm ph}$; its lower boundary is $\Delta t_{\rm ph,min}$, which is calculated as median($\{\delta t_{p,i}\})$. The rationale is that the duration of a phase bin cannot be shorter than the typical uncertainty of the peak times.
    \item $\Delta t_{\rm bulk}$ is the shortest contiguous time interval comprising 50\% of the peak times.
\end{itemize}

In the following, we detail how each of the quantities introduced above was calculated or why it was fixed to a specific value.

$N_{\rm cyc,min}$ determines the longest trial period and was set to $8$. Admitting values $N_{\rm cyc,min}<8$ would imply a low-quality factor, $Q\le N_{\rm cyc}$, for the putative periodic signal and, consequently, a weak claim for periodicity. In addition, we also imposed the condition $P\le \Delta t_{\rm bulk}$: the reason is that we do not want to consider the case of a period longer than the time interval that includes a sizeable fraction of peak times (50\%, specifically), since it is clearly irrelevant and could contribute to increasing the rate of spurious periodic candidates.

On the opposite side, the shortest trial period is given by $N_{\rm cyc,max}$: this was calculated as the closest integer to $T/(N_{\rm ph,min} \Delta t_{\rm ph,min})$. In fact, the shortest explorable period was obtained with the product of the minimum number of phase bins multiplied by the shortest phase bin duration. $N_{\rm cyc}$ assumes all integer values between the two boundaries for the same reason that, in a Fourier PDS extracted over a time window, $T$, all frequencies that are evenly spaced by $1/T$ up to the Nyquist one are independent.

Given $P$, we tried a range of values for the number of phases, $N_{\rm ph}$: the minimum value was fixed to $N_{\rm ph,min}=5$, since a lower value would correspond to a rough sensitivity and large number of spurious candidates. $N_{\rm ph,max}$ was calculated as the maximum between $N_{\rm ph,min}$ and the rounded integer of $P/\Delta t_{\rm ph,min}$: again, the reason was that we rejected the case $\Delta t_{\rm ph} < \Delta t_{\rm ph,min}$.
Thus far, for a given trial, $P$, or equivalently $N_{\rm cyc}$, we determined the array of possible values for the number of phases, $N_{\rm ph}$: the array of values for $N_{\rm ph}$ was chosen as $N_{\rm ph,min} 2^n$
with $n=0,1,\ldots, [\log_2{(N_{\rm ph,max}/N_{\rm ph,min})}]$; that is, doubling the number of phases until it exceeded the upper boundary.

For any given pair of values for $(N_{\rm cyc}, N_{\rm ph})$, all the remaining parameters were automatically determined. At this stage, given a GRB and having defined the range for both quantities, for any combination of $(N_{\rm cyc}, N_{\rm ph})$ we folded the profile accordingly and counted how many peaks fell in each phase bin. We let $c_i$ be the number of peaks falling in phase bin $i$. We assigned a relative likelihood to the pair $(N_{\rm cyc}, N_{\rm ph})$ by means of the probability mass function of the multinomial distribution, which was defined as
\begin{equation}
    f({\bf c},N_p,{\bf p})\ =\ \frac{N_p!}{c_1! c_2!\ldots\ c_{N_{\rm ph}}!}\ p_1^{c_1} p_2^{c_2}\ldots p_{N_{\rm ph}}^{c_{N_{\rm ph}}}\ =\ N_p!\ \Pi_{i=1}^{N_{\rm ph}} \frac{p_i^{c_i}}{c_i!}\;,
    \label{eq:multinom}
\end{equation}
where $\sum_i c_i = N_p$ and ${\bf p}$ are the probabilities of the corresponding phases. In our case, all phases are equivalent, so we took $p_i=1/N_{\rm ph}$ for all $i$.
To properly account for the probability space in the null hypothesis of no recurrent behaviour in the sequence of peaks, for each GRB we simulated a suite of $N_{\rm sim}=10^4$ random sequences of $N_p$ non-periodic peak times. Each sequence then underwent the same analysis in terms of the multinomial probability as in Eq.~(\ref{eq:multinom}) and the resulting distribution was used as a reference for the real sequence. 

The $N_{\rm sim}$ samples of peak times should reproduce the same non-stationary character of the real peaks, randomly distributed and under the null hypothesis of no periodic signal. This is a fundamental aspect, missing which we would test a null hypothesis of no periodic recurrences in peaks, which however would not reproduce a key property of real GRB LCs: their non-stationary character. In fact, for a generic GRB LC, peaks are not observed with the same rate over time, but it is known that they tend to cluster around periods of activity, which are occasionally interspersed with quiescent times. While this simple property does not necessarily imply a bias either in favour of or against possible periodic recurrences of the peaks, it surely alters in a non-trivial way the simplistic null hypothesis of a constant probability per unit time for a peak to occur. After all, the importance of a credible null hypothesis should not be underestimated, since its rejection is then used to claim for periodic recurrences in the peak times.

We let ${\bf t}^{(k)}_{p} = \{t^{(k)}_{p,i}\}$ $(i=1,\ldots,N_p)$ be the $k$-th sequence $(k=1,\ldots,N_{\rm sim})$ of $N_p$ randomly simulated and non-periodic peak times.  In Sect.~\ref{sec:non-stat}, we describe how we simulated ${\bf t}^{(k)}_p$ for $k=1,\ldots,N_{\rm sim}$, in a way that accounts for the non-stationarity of GRBs.

For any given pair $(N_{\rm cyc}, N_{\rm ph})$, each sequence, ${\bf t}^{(k)}_p$, was folded into a histogram of counts per phase bin, which we denote as $\pmb{\zeta}^{(k)} = \{\zeta^{(k)}_{i}\}$ $(i=1,\ldots,N_{\rm ph})$, to be compared with the analogous histogram, ${\bf c} = \{c_i\}$, corresponding to the real sequence. 

Insofar as the simulated sequences of peak times reproduce the non-stationary character of the real data, the likelihood assigned to a generic $k$-th sample for a given $(N_{\rm cyc}$, $N_{\rm ph}$) is given by Eq.~(\ref{eq:multinom_cond_fake}), which is equivalent to Eq.~(\ref{eq:multinom}) properly renormalised by the summed probability of all simulated samples, $P_{\rm sum}(N_{\rm cyc}, N_{\rm ph})$,
\begin{equation}
    f_{\rm ren}(\pmb{\zeta}^{(k)},N_p,{\bf p}) =\ \frac{1}{P_{\rm sum}(N_{\rm cyc}, N_{\rm ph})}\ N_p!\ \Pi_{i=1}^{N_{\rm ph}} \frac{p_i^{\zeta^{(k)}_i}}{\zeta^{(k)}_i!}\;,
    \label{eq:multinom_cond_fake}
\end{equation}
where $P_{\rm sum}(N_{\rm cyc}, N_{\rm ph})$ is defined as
\begin{equation}
    P_{\rm sum}(N_{\rm cyc}, N_{\rm ph})\ =\ \sum_{k=1}^{N_{\rm sim}}\ f(\pmb{\zeta}^{(k)},N_p,{\bf p})\;.
    \label{eq:multinom_cond_sum}
\end{equation}
This is evaluated as the sum over all the $N_{\rm sim}$ sets, which cover the parameter subspace of peak times compatibly with the LC of every given GRB, such that the sum over $k$ of Eq.~(\ref{eq:multinom_cond_fake}) is 1. 

The convenience of renormalisation as in Eqs.~(\ref{eq:multinom_cond_fake} and \ref{eq:multinom_cond_sum}) is due to the fact that the typical values calculated with Eq.~(\ref{eq:multinom}) change remarkably for different values for $(N_{\rm cyc}, N_{\rm ph})$. Without the renormalisation term of Eq.~(\ref{eq:multinom_cond_sum}), we noticed that the method, whose sensitivity was established through the calibration procedure described in Sect.~\ref{sec:cal}, became almost blind, due to the biased comparison of different values of $(N_{\rm cyc}, N_{\rm ph})$.

Analogously to Eq.~(\ref{eq:multinom_cond_fake}), which applies to the synthetic sequences, we calculated the same likelihood for the real sequence:
\begin{equation}
    {\cal L}(N_{\rm cyc}, N_{\rm ph})\ =\ f_{\rm ren}({\bf c},N_p,{\bf p}) =\ \frac{1}{P_{\rm sum}(N_{\rm cyc}, N_{\rm ph})}\ N_p!\ \Pi_{i=1}^{N_{\rm ph}} \frac{p_i^{c_i}}{c_i!}\;.
    \label{eq:multinom_cond_real}
\end{equation}
Finally, the best candidate to reject the null hypothesis of no periodic recurrence in real peak times corresponds to the pair $(N_{\rm cyc,best}, N_{\rm ph,best})$ that minimises the likelihood associated with the null hypothesis, which is Eq.~(\ref{eq:multinom_cond_real}):
\begin{equation}
    {\cal L}(N_{\rm cyc,best}, N_{\rm ph,best})\ =\ \min_{N_{\rm cyc}, N_{\rm ph}}\ {\cal L}(N_{\rm cyc}, N_{\rm ph})\;.
    \label{eq:multinom_cond_real_best}
\end{equation}

The best period candidate is therefore $P_{\rm best} = T/N_{\rm cyc,best}$.
The final statistical significance of Eq.~(\ref{eq:multinom_cond_real_best}), which we define as the `p value', since it accounts for the complex multi-trial correlation involved between different trials, and non-stationarity, is given by comparing the result of Eq.~(\ref{eq:multinom_cond_real_best}) with the distribution of $N_{\rm sim}$ analogous values obtained for each of the simulated sequences of non-periodic peaks $\{{\bf t}^{(k)}\}$: the best candidate of each synthetic sequence is given by minimising Eq.~(\ref{eq:multinom_cond_fake}) over all pairs $(N_{\rm cyc}, N_{\rm ph})$ in the very same way as for the real sequence. Summing up, the p value is the probability of a given GRB having no periodic recurrent peak times.

\begin{figure}[!h]
   \centering
   \includegraphics[width=0.5\textwidth]{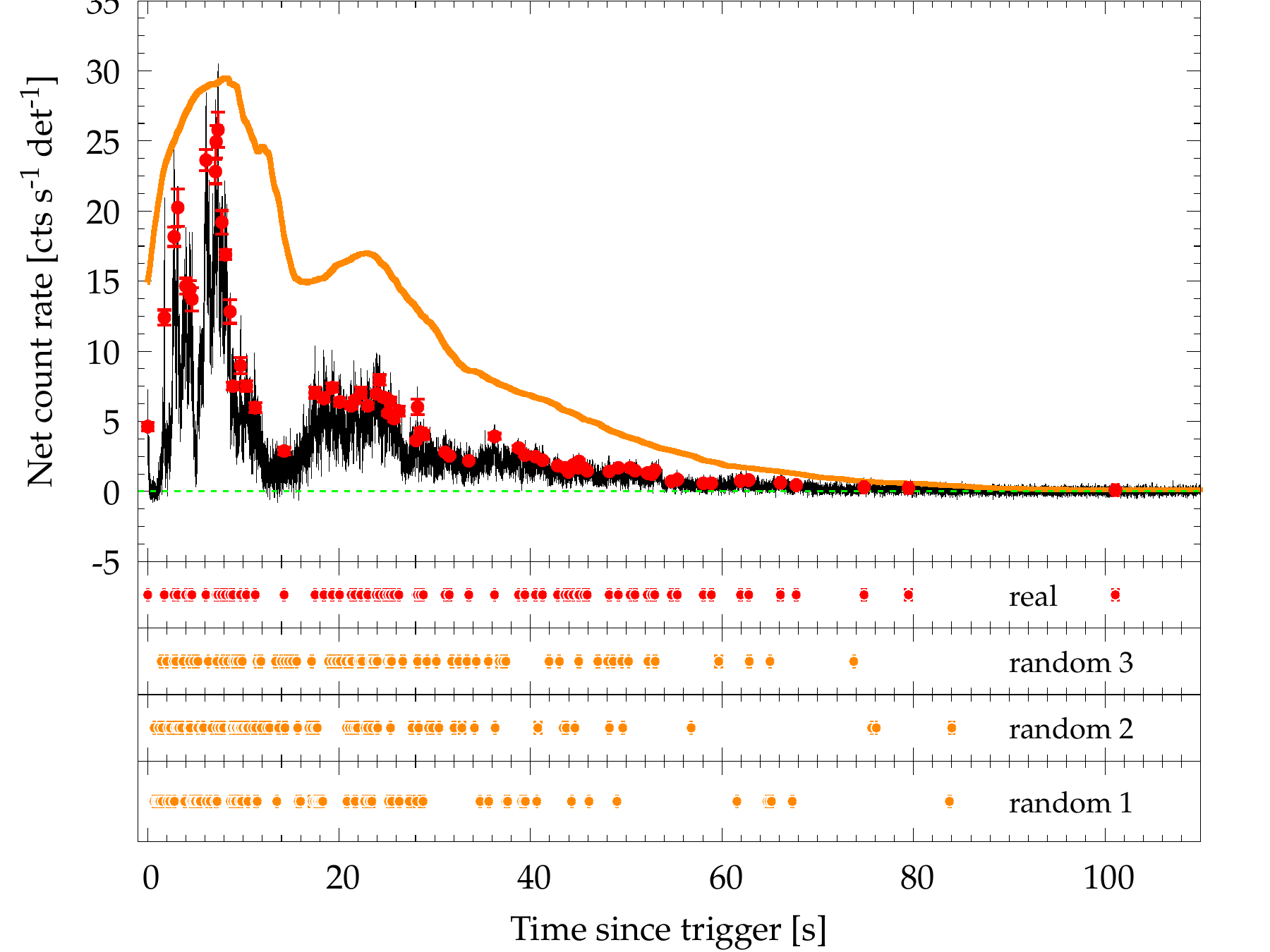}
    \caption{{\it Top panel}: LC of GRB\,211211A observed by {\it Swift}/BAT in the 15-150~keV passband as an illustrative example. The bin time is 4~ms. The identified peaks are highlighted with red circles. The solid orange line shows the probability density function, $\Psi(t)$ (Sect.~\ref{sec:non-stat}), with arbitrary normalisation, assumed to generate fake peaks, and was obtained from smoothing the original LC with a boxcar window of $13.5$~s to account for the non-stationary nature of GRB profiles (see Sect.~\ref{sec:non-stat}). {\it Small bottom panels}: Three different examples (orange circles) of random sequences of peak times, $\pmb{\zeta}^{(k)}$, generated using $\Psi(t)$ shown in the top panel; the red circles show the real peak times, also shown in the top panel.}
    \label{fig:211211A}
\end{figure}
%
%-----------------------------------------
\subsection{Non-stationarity of GRB profiles}
\label{sec:non-stat}
%-----------------------------------------
As it is well known in the Fourier space, the presence of a non-stationary signal makes the search for periodic or quasi-periodic features challenging, because the null hypothesis of no periodicity to be tested can hardly be recognised, to the same extent that searching for something in an unknown environment is more challenging than searching for it in a known one. The analogous issue in our case manifests itself in the way synthetic peaks are generated as a function of time. In Sect.~\ref{sec:ana} we anticipated that, for each GRB, the $N_{\rm sim}$ synthetic sequences of $N_p$ peak times are generated so that they reproduce the non-stationary character of the real LC. Here, we describe specifically how this was done in practice.

We started from the general property of multi-peaked GRB LCs, in which peaks tend to cluster around periods of activity. The average count rate during these active windows is typically higher than in other time intervals for the obvious reason that the presence itself of multiple peaks within a given time interval, contributes to enhance the resulting flux.
This property suggested that we should adopt the GRB time profile itself to describe the probability per unit of time of a peak occurring, which we denote as $\Psi(t)$. Clearly, $\Psi(t)$ should not follow the ebbs and flows of the real LC in detail; otherwise, it would preferably simulate peaks at the exact times of the real ones, at the risk of blinding the search for genuine periodic recurrences.

The function, $\Psi(t)$, has to be the result of some smoothing of the 
original LC. The process of smoothing a time series when searching for 
periodicity is called `windowing', and its goal is to reduce the spectral 
leakage due to the finite duration of the observation (in our specific case, 
the finite duration of GRBs).  There are several different types of window functions that can be applied, depending on the peculiarity of the signal. When preserving frequency resolution is a primary concern, 
then, for example, the Kaiser-Bessel is a good choice. On the other hand, when the primary concern is the amplitudes of the frequencies -- the ability to distinguish how many peak times fall within a certain time interval -- then a flat-top (boxcar) window is a good choice. A reasonable trade-off between frequency and amplitude accuracy is offered by the Hann window.\footnote{We used the function {\tt hann} from the Python package {\tt scipy.signal}.} For this reason, we convolved the GRB LCs with both the boxcar and the Hann windows, in order to explore the extent to which our results depend on the particular choice of window. As we detail below, the analysis confirmed the independence of the results from the window. In the following, we show the results obtained with the boxcar window, unless specified differently.

After several attempts, the optimal duration of the boxcar window was identified in the shortest contiguous interval that collected a given fraction (we chose 50\%) of all the peak times. This choice turned out to be effective: on the one hand, the exact information on the sequence of a sizeable fraction of peaks is lost; on the other hand the overall profile is reproduced.

Figure~\ref{fig:211211A} shows the example of the famous GRB\,211211A as seen with {\it Swift}/BAT in the 15-150 keV passband. Along with the original 4-ms time profile, also shown are the peaks identified with {\sc mepsa} as well as the function, $\Psi(t)$ (solid orange line), which resulted from the convolution of the original LC with a boxcar window of $13.5$~s. Clearly, $\Psi(t)$ reproduces the overall behaviour over medium and long timescales, while it loses any information on most waiting times\footnote{Time intervals between adjacent peaks.} between adjacent peaks.

Once $\Psi(t)$ was calculated, for each synthetic sequence we simulated one peak at a time using $\Psi(t)$, conveniently renormalised, as a probability per unit time. To each of the fake peak times, $t^{(f)}_{p,i}$, we assigned the uncertainty, $\delta t_{p,i}$, of the corresponding real peak time and accepted it only if there was no overlap between the interval, $t^{(f)}_{p,i} \pm \delta t_{p,i}$ and the analogous intervals of the other fake peaks. This is an important requirement, which ensures the same separability as the real peaks.
Upon the acceptance of the $N_{p}$-th peak, a given synthetic sequence was complete.

%-----------------------------------------
\subsection{Study of the technique sensitivity}
\label{sec:cal}
%-----------------------------------------
Before we analysed the real GRBs, we characterised our technique as follows.
For each GRB of the five data sets, we simulated a fake sequence of peak times that had as many peaks as the real one.
For each sequence, we randomly selected a period within the interval $[P_{\rm min}, P_{\rm max}]$, assuming a uniform distribution in $\log{P}$; we denote this fake period as $P_f$. Within the time interval covered by $\Psi(t)$, we randomly selected $N_p$ times out of a sequence of $P_f$-periodic times, as follows: for any of these periodic times, $t^{(p)}_{p,i}$, we assumed a Bernoulli distribution with a probability, $\Psi(t^{(p)}_{p,i})$, of being taken, going on, and looping until $N_p$ fake and periodic peak times were accepted. Upon acceptance of each peak time, we verified the same separability condition based on the uncertainties, $\delta t_{p,i}$, affecting the real peak times as in Sect.~\ref{sec:non-stat}. Lastly, each accepted fake peak time, $t^{(p)}_{p,i}$, was randomly scattered assuming a uniform distribution in the interval $t^{(p)}_{p,i}\pm \delta t_{p,i}$ to incorporate the impact of the measurement errors that affect real peaks. 

Finally, we generated three more fake sequences of $N_p$ peaks, but with a different fraction, $\xi$, of periodic peaks, with $\xi=0.75$, $\xi=0.5$, and $\xi=0.25$, respectively (integer parts), with the remaining peaks chosen randomly. Therefore, calling $N_{\rm per}$ the number of periodic peaks, it is $N_{\rm per} = [\xi N_p]$. While the first sequence has 100\% periodic peaks, the next three ones reproduce the possible case of periodic recurrent peaks being mixed with other unrelated peaks.

At this point, for each GRB, each of the four sequences of fake periodic peaks was treated as a real sequence and consequently went through the same procedure outlined in Sect.~\ref{sec:ana}. For each case, we extracted the best period candidate, $P_{\rm best}$, along with its p value, and compared it with the correct $P_f$.
\begin{figure}
   \includegraphics[width=0.5\textwidth]{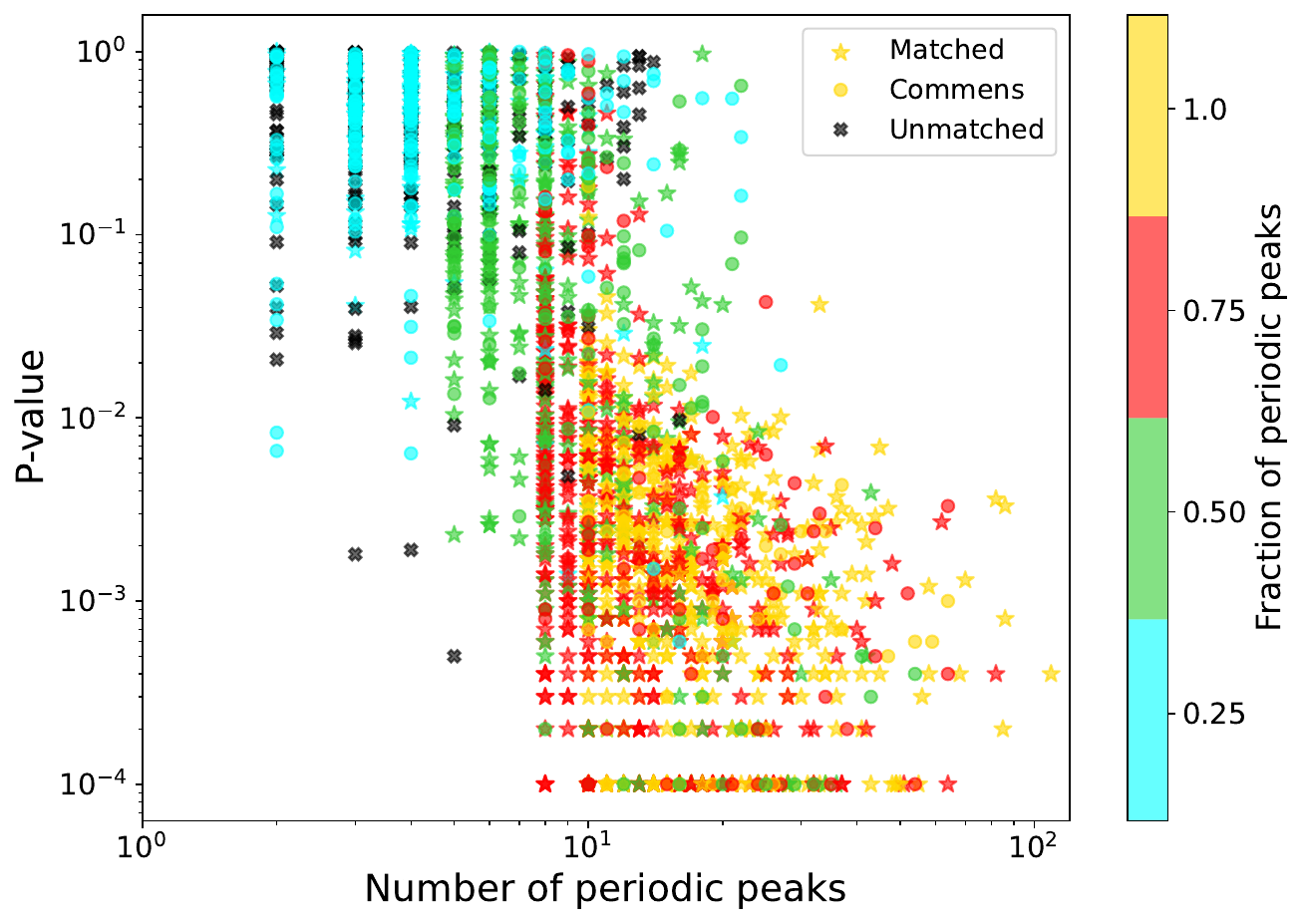}
    \caption{P values of fake GRBs as a function of the number of periodic peaks for the three different cases: the simulated period is (i) correctly matched by the procedure (star), (ii) commensurate with the best candidate, mostly by a factor of 2 or 3 (circle), (iii) is not correctly identified by the procedure (black cross). Colours reflect the fraction of periodic peaks over the total number of peaks within each single simulated GRB.}
    \label{fig:EffectPV_fake}
\end{figure}

Figure~\ref{fig:EffectPV_fake} shows the scatter plot for all of the GRBs of the total sample in the $N_{\rm per}$--p-value plane. The colour code corresponds to the fraction, $\xi$, of periodic peaks, whereas different symbols tell whether (i) it was a correct identification, $P_f\simeq P_{\rm best}$ (star), (ii) or $P_f\ne P_{\rm best}$, but $P_f$ was commensurate with $P_{\rm best}$ (circle), or (iii) or remained unidentified (black cross). The points with p values $\lesssim 5\times10^{-4}$ should be taken as either rough or upper limits, as they resulted from $N_{\rm sim}=10^4$ simulations.

Overall, correct identifications amount to 61\%, commensurate ones to 25\%, and the unmatched cases to 13\%. Interestingly, 60\% of the correct identifications are characterised by a p value $< 0.01$, $\xi\ge0.75$, and $N_{\rm per}\gtrsim 8$ (yellow/red stars). The same combination of properties among the commensurate cases results in a fraction of 19\%, which is lower than for the correct identifications: the reason is that a sizeable fraction of commensurate identifications are due to fewer ($\lesssim 8$) periodic peaks and/or high ($\gtrsim 0.1$) p values.

Taking the candidates with p value $<0.01$, $N_{\rm per}\gtrsim 8$ periodic peaks, and $\xi\ge 0.75$, which is equivalent to $N_p\gtrsim N_{\rm per}/\xi \simeq 10$, all of them are either correctly identified (88.5\%) or commensurate candidates (11.3\%), neglecting 0.2\% that are unmatched. The p value of this subsample has a median value of $1.8\times 10^{-3}$ and a 90 percentile of $1.9\times 10^{-2}$. For the commensurate identifications, the ratio between $P_f$ and $P_{\rm best}$ is either 2 or 3 in 91\% of cases.

\begin{figure}
   \includegraphics[width=0.5\textwidth]{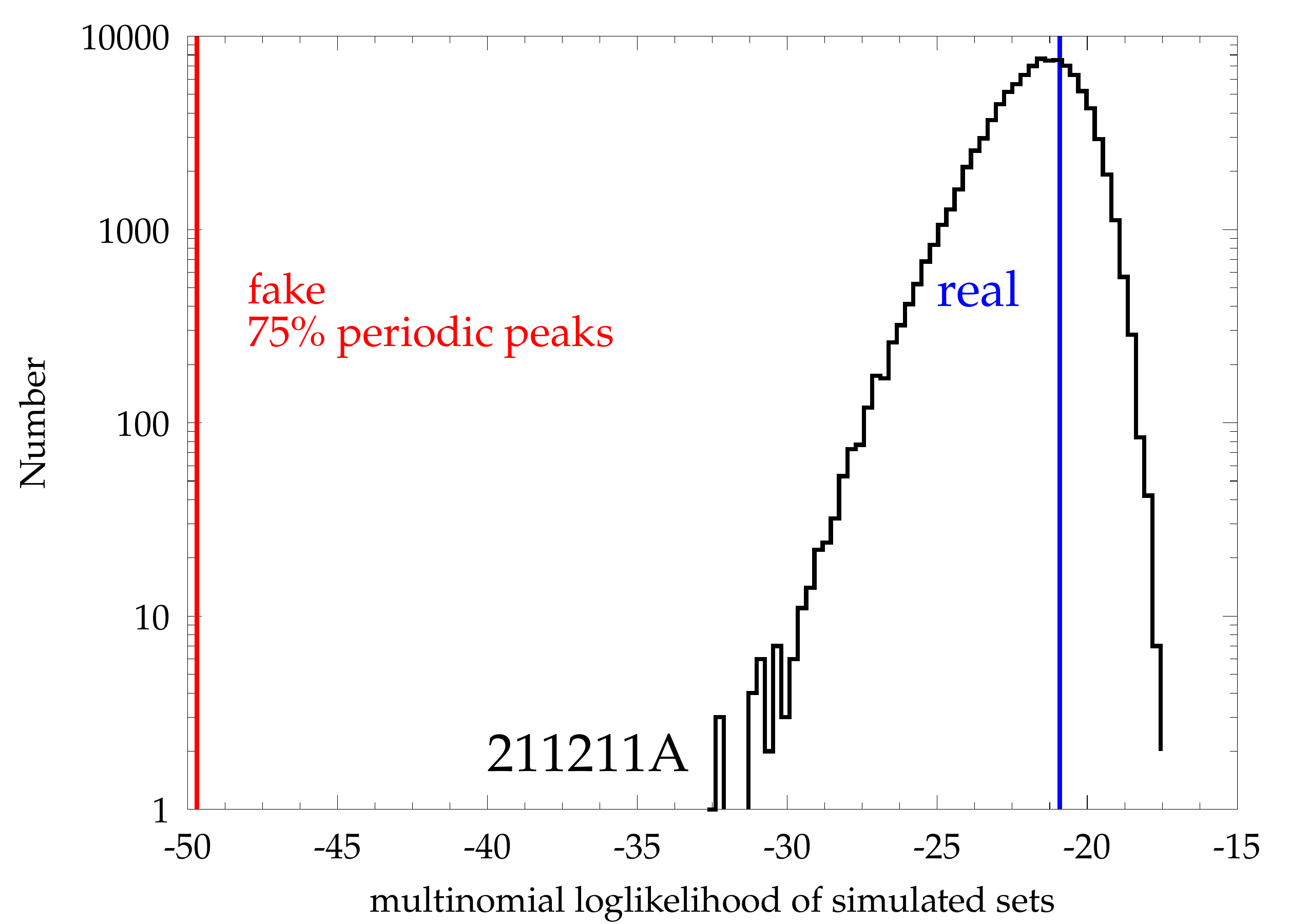}
    \caption{Distribution of the multinomial likelihood values calculated for each of the $10^5$ simulated sets of peak times under the assumption of no periodic occurrences, in the case of GRB\,211211A, also shown in Fig.~\ref{fig:211211A}. The vertical blue line shows the corresponding value for the real data, showing that the no-periodic occurences assumption cannot be rejected. Instead, the vertical red line shows the analogous result for a 211211A-like GRB sharing the same number of peaks and non-stationary profile, but with 75\% periodic peaks. The p value is clearly $\ll 10^{-5}$. }
    \label{fig:logsim}
\end{figure}

Figure~\ref{fig:logsim} shows the resulting distribution of $10^5$ multinomial likelihood values obtained from as many random sequences of peak times in the case of GRB\,211211A ({\it Swift}/BAT data). While the real sequence of peaks (vertical blue line) appears to be fully compatible with the assumption of no periodic recurrences, the simulated case of 75\% periodic peaks has a p value that is $\ll 10^{-5}$. An accurate determination of its value would require much more intensive calculations, which, within the scope of the present work, would only be worth it in the event of a real and equally significant case. In this example, we increased by one decade the number of simulations just to showcase the sensitivity of the method for a GRB featuring a lot of peaks (72 for GRB\,211211A), three quarters of which would occur periodically.

By relaxing the threshold on $\xi$ to $\xi\ge 0.5$, so including the green points in Fig.~\ref{fig:EffectPV_fake}, these present a wide range of p values: this means that, whenever the fraction of periodic peaks is 50\%, a correct (or commensurate) identification with a p value $<0.01$ is possible, but it is not granted. For all these simulations, the smoothing function, $\Psi(t)$, was obtained with a boxcar window (Sect.~\ref{sec:non-stat}). We carried out the analogous set of simulations, replacing the boxcar with the Hann filter, and the results do not change in essence, obtaining the same indications that were derived above for a boxcar. Summing up, as long as a GRB features $N_{\rm per}\gtrsim 8$ periodic recurrent peaks and they make up at least three quarters of all peaks, our procedure is able to identify the recurrence in 50\% (90\%) of cases with a p value $< 1.8\times 10^{-3}$ ($< 1.9\times 10^{-2}$).

%%%%%%%%%%%%%%%%%%%%%%%%%%%%%%%%%%%%%%%%%%
\section{Results}
\label{sec:res}
%%%%%%%%%%%%%%%%%%%%%%%%%%%%%%%%%%%%%%%%%%
%
\begin{figure}
   \includegraphics[width=0.5\textwidth]{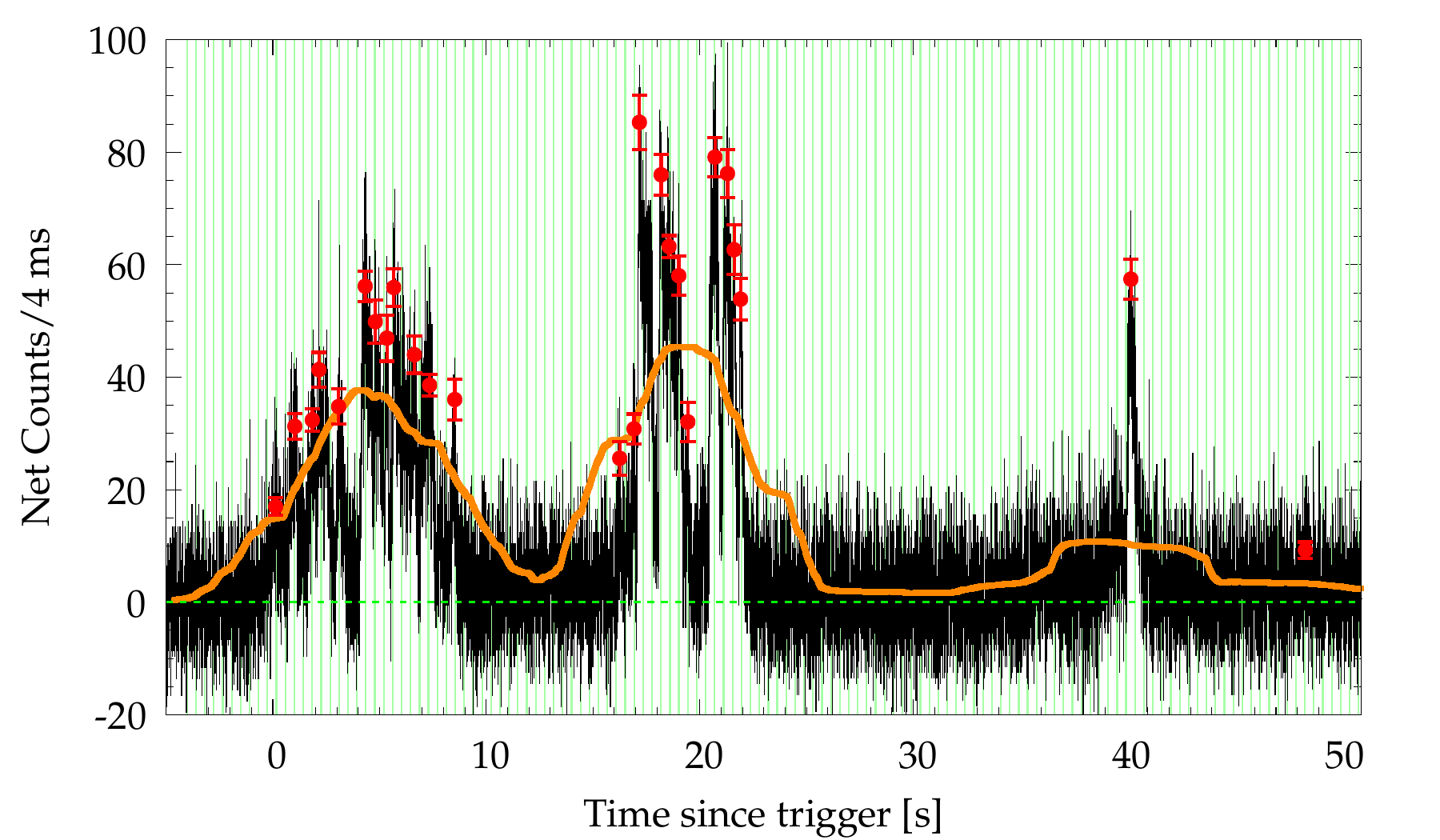}
   \includegraphics[width=0.5\textwidth]{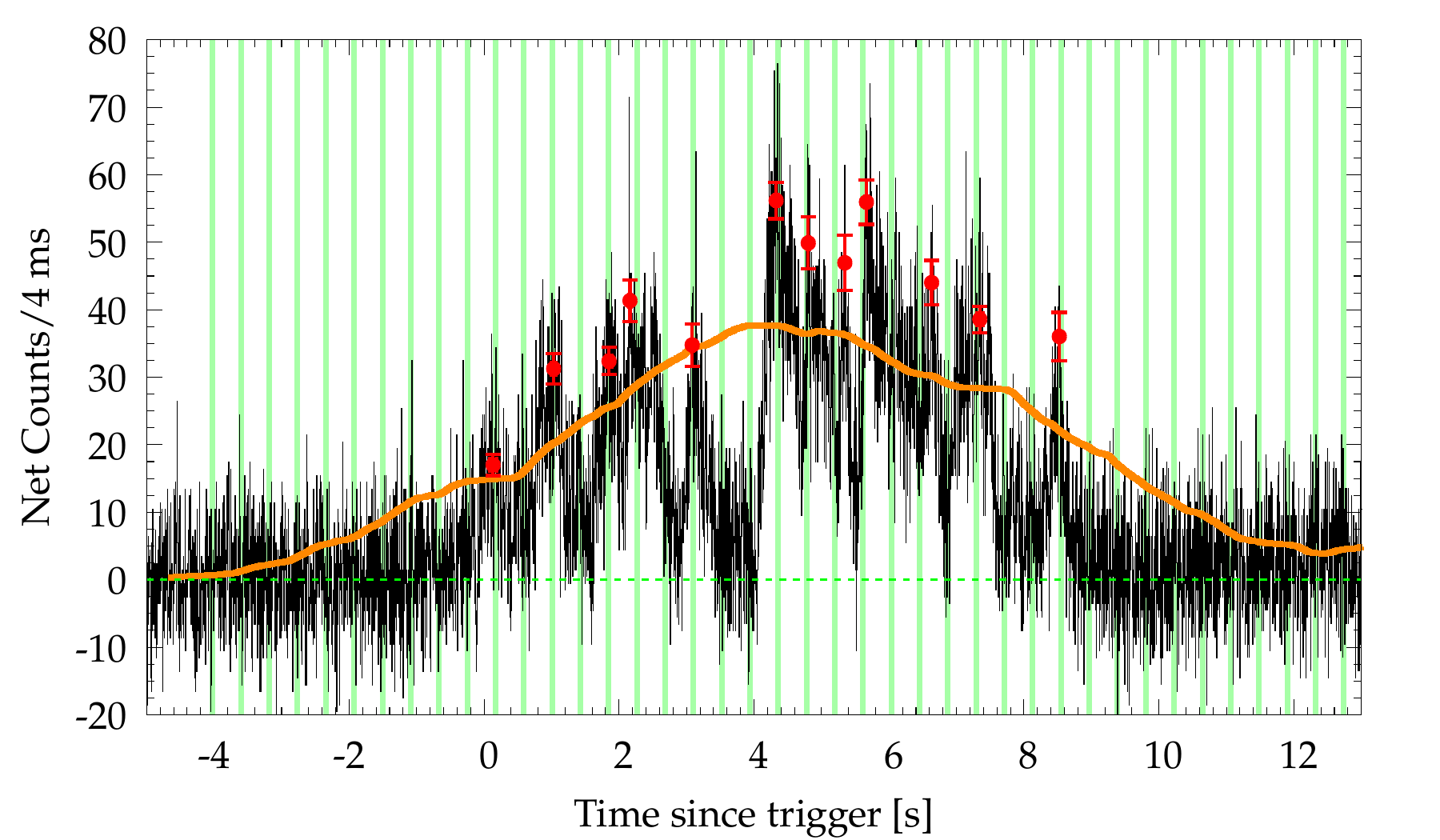}
   \includegraphics[width=0.5\textwidth]{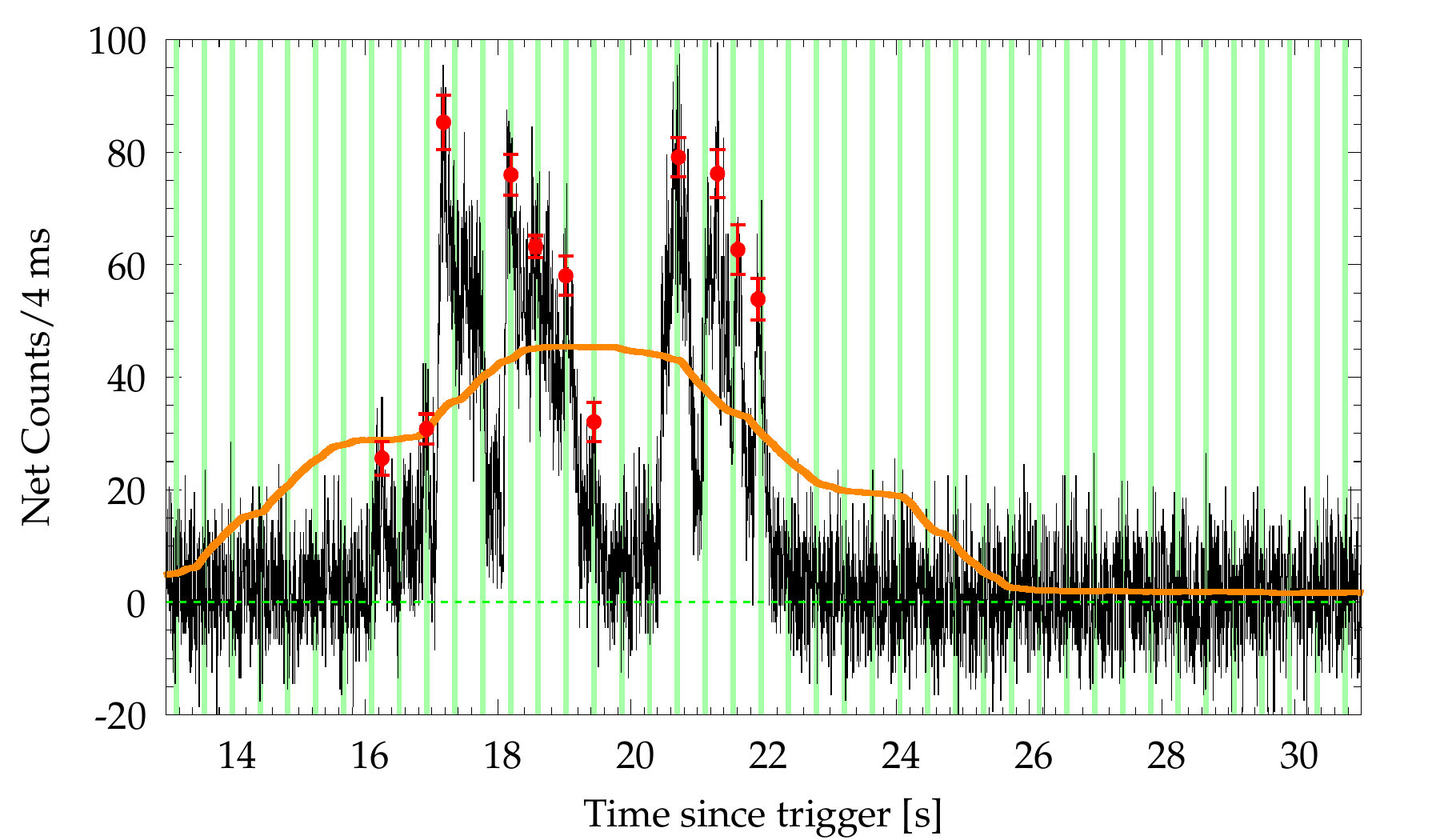}
   \includegraphics[width=0.5\textwidth]{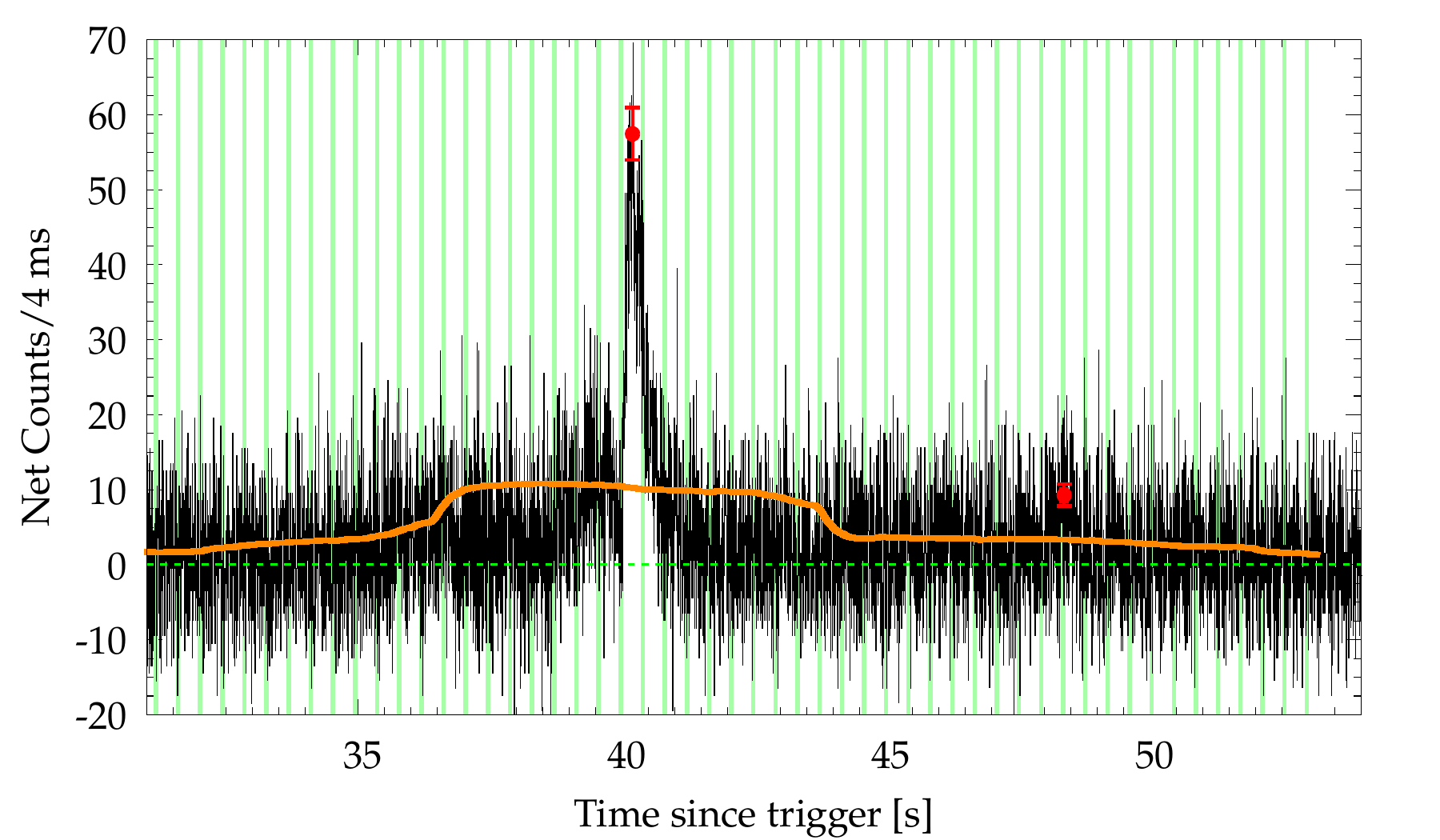}
    \caption{LC of the best candidate. Top: {\it Fermi}/GBM bn170207906 is the best candidate for a periodic recurrence of peaks: 17 out of a total of 25 peaks (red circles) belong to the same phase bin (vertical green regions) out of 5 phase bins, corresponding to a period of $0.419$~s. The p value is $3\times10^{-4}$. The solid orange line is the probability per unit of time, $\Psi(t)$ (arbitrary normalisation), used to generate fake peaks to estimate the statistical significance. The panels from second top to bottom are close-in views of three different portions of the total LC shown in the top panel.}
    \label{fig:bestLC}
\end{figure}
We applied the procedure described in Sect.~\ref{sec:ana} to the 555 GRBs in the total sample. The GRB with the lowest p value of $3\times10^{-4}$ as calculated 
in Eq.~(\ref{eq:multinom_cond_real_best}) is {\it Fermi}/GBM bn170207906: this GRB has 25 peaks, 17 of which ($\sim 70$\%) are found in the same phase with a period of $0.419$~s split into five~phase bins. Figure~\ref{fig:bestLC} displays its LC along with its peaks and with the putative periodic phase bins highlighted. Based on the characterisation of the technique reported in Sect.~\ref{sec:non-stat}, it is tempting to consider this GRB as a convincing case for a periodic recurrence in peaks. However, correcting for the overall multi-trial due to searching 555 GRBs degrades the statistical significance of the best candidate to a mere 17\%. The three lowest p values are in the range of $0.3$--$1.1\times10^{-3}$: the Poisson probability of having $\ge 3$ out of 555 candidates with comparably low or lower p values is 2.4\%, which is not small enough to claim that there is a significant presence of periodic GRBs. Tables 1 to 5 report the results for the best candidate of each GRB in the {\it Swift}, BATSE, {\it Fermi}/GBM, {\it Insight/HXMT}, and {\it Beppo}SAX samples, respectively.
We obtained similar results -- that is, no significant candidates -- when we assumed the Hann instead of the boxcar window, to model the non-stationarity (Sect.~\ref{sec:non-stat}).

\begin{figure}
   \includegraphics[width=0.47\textwidth]{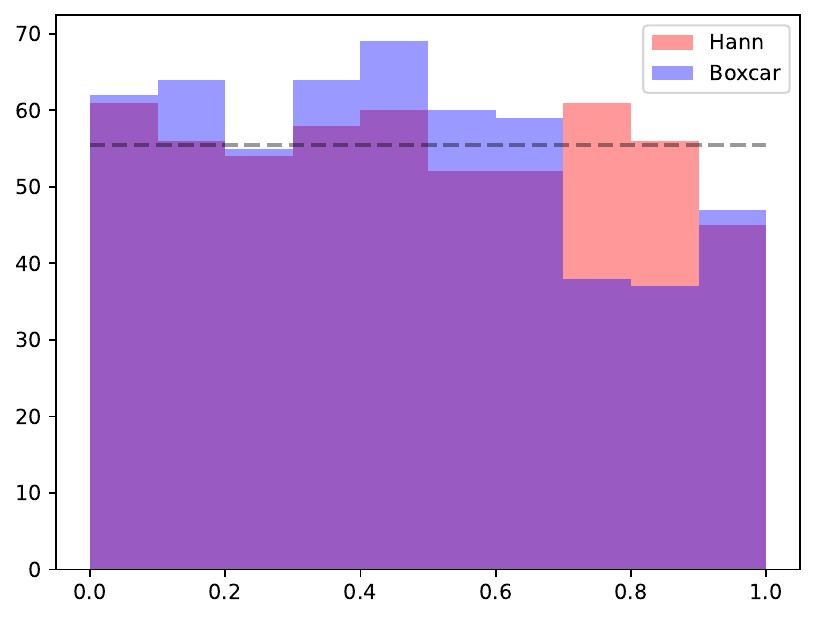}
    \caption{Distributions of p values (Eq.~\ref{eq:multinom_cond_real_best}) for all 555 GRBs in the total sample, using either a boxcar or a Hann filter to model the non-stationarity of GRB LCs. The dashed line shows the average value that is expected for uniform distribution.} 
    \label{fig:KS_Unif}
\end{figure}

Looking at the distribution of 555 p values (Fig.~\ref{fig:KS_Unif}), one for each GRB, we wondered whether this is overall compatible with a uniform distribution in the $[0,1]$ range, as one would expect if there were a systematic and general lack of evidence of periodic recurrent peaks. A Kolmogorov-Smirnov test yields a p value of $0.06$\% for the boxcar and 61\% for the Hann windows. Hence, no robust evidence is found that rejects the null hypothesis of no  GRBs with periodic recurrences in peak times, especially in light of the significance of the best candidate. Lastly, the distribution of the trial periods that have been examined for each of the five GRB samples shows that the range spans from a few~$\times\,0.1$ to a few~$\times\,10$~s, covering two decades.
%

%%%%%%%%%%%%%%%%%%%%%%%%%%%%%%%%%%%%%%%%%%
\section{Discussion and conclusions}
\label{sec:disc_conc}
%%%%%%%%%%%%%%%%%%%%%%%%%%%%%%%%%%%%%%%%%%
We devised a new method of searching for (quasi-)periodic components in long-duration GRBs featuring at least ten peaks, after reformulating the problem. Compared with the traditional techniques that have so far been adopted in the literature to tackle the same problem, both in frequency and in time domains, our technique just focuses on the times of the peaks, once these are reliably identified. Our method reduces the information stored in a GRB time series to a point process, such as the one made up by the peak times. As such, it is not meant to replace more traditional techniques, but at the cost of losing some information it simplifies the problem, alleviating some of the pitfalls that surround the other techniques, such as red noise and non-stationarity. Its robustness also relies on very simple and basic requirements, which are (i) the identification of peaks through a reliable tool like {\sc mepsa}, and (ii) an effective procedure to account for the non-stationarity of GRB LCs, which does not depend on the kind of smoothing.

The idea of focusing on the sequence of events in time is often used in other fields; for instance, the discovery that some FRB sources are periodic repeaters just adopted the same approach \citep{CHIME20a}. In general, every phenomenon that can be described in terms of a point process can naturally be addressed in this way.

Interestingly, like any folding technique, this method is sensitive to periodically recurrent events, even when some expected periods of activity turn out to be quiet, as is indeed the case for periodic FRB repeaters. Another advantage of this method is its robustness against the event with unidentified peaks: missing a peak would just make the analysis slightly less sensitive, but there is no pile-up effect, which instead would affect the study of the distribution of waiting times.

Having established that this method simplifies the search for periodic recurrences in GRB LCs, the question is whether it may detect the same kinds of oscillations that have been discussed in the GRB literature so far. Surely, the high-frequency and especially `transient' oscillations are beyond the scope of our method, unless they feature several statistically significant ms-long peaks, which is never the case with past and current instrumentation and data sets. But persistent oscillations (at least with 8--10 cycles) in the range from a few $\times$$0.1$ to a few $\times$$10$~s can be detected with $\gtrsim 3\sigma$ confidence, provided that the number of recurrent peaks dominates ($\gtrsim 75$\%) the overall number of peaks within a given GRB LC, as we proved in Sect.~\ref{sec:cal}.

Overall, our analysis of 555 multi-peaked GRBs from five different experiments does not come up with any very strong candidate of periodic recurrences, once all the kinds of involved multi-trials are properly accounted for. In addition, we found no robust hint that, for a sizeable fraction of GRBs, the majority of peaks occur in a deterministic way in time (as opposed to randomly), once the non-stationarity of the phenomenon is considered. These conclusions hold true at least for the range of values, which cover the spin periods of most Galactic magnetars discovered to date. Clearly, this cannot rule out ms-magnetars being GRB engines \citep{Usov92,Duncan92,Zhang01,Metzger11} for many reasons. Firstly, at the time of the GRB, their estimated initial spin period is in the millisecond range \citep{Dallosso11}. Secondly, even if the engine releases energy with a periodic recurrence over time, the unknown dissipation mechanism into gamma-rays could either wash out or significantly degrade any imprinted coherence. For instance, the internal shock (IS) model \citep{Rees94} predicts a one-to-one relation between peaks and source activity \citep{Kobayashi97}: should the source emit the different shells with regular separations, the distribution of Lorentz factors, which contribute to determining the collision times, could modulate and wash out the imprinted periodicity.

In cases of special interest such as GRB\,211211A, our explored range of periods is $[0.36, 15.15]$~s, so our method cannot test the existence of four oscillations with $P\sim 0.05$~s over a $0.2$-s long interval \citep{Xiao24,Chirenti24}. Yet, when the number of oscillations is so small, especially in the light of the various potential issues mentioned above (Sect.~\ref{sec:intro}), a strong word of caution is needed.

In conclusion, we found no case of long, multi-peaked GRBs that display a statistically significant periodic recurrence in peak times, even admitting the possibility that a fraction up to $\sim 25$\% of all peaks occur at random (or unrelated) times. Consequently, either the intermittent character that characterises the energy release is inherently non-periodic, or the dissipation mechanism into gamma-rays washes out any originally imprinted coherence. 

\section*{Data availability}
Tables~1 to 5 are only available in electronic form at the CDS via anonymous ftp to cdsarc.u-strasbg.fr (130.79.128.5) or via http://cdsweb.u-strasbg.fr/cgi-bin/qcat?J/A+A/.

\begin{acknowledgements}
We are grateful to the anonymous reviewer for their insightful feedback which helped us to improve the manuscript. L.F. acknowledges support from the AHEAD-2020 Project grant agreement 871158 of the European Union’s Horizon 2020 Program.
\end{acknowledgements}

\end{document}